\documentclass[12pt]{article}

\newcommand{\simonius}{\author{\normalsize Markus Simonius\\
\normalsize
Institut f\"ur Teilchenphysik, Eidgen\"ossische Technische Hochschule,\\
\normalsize
CH-8093 Z\"urich, Switzerland}}

\newcommand{\qed}{\nolinebreak\rule[-0.5ex]{0.4em}{2.1ex}}

\newtheorem{thm}{Theorem}
\newtheorem{cor}[thm]{Corollary}

\begin{document}

\title{Measurement in Quantum Mechanics: \\ From Probabilities to
Objective Events 
\footnote{Published in {\em Helvetica Physica Acta} 66 (1993) 721.}
}
\simonius
\date{}
\maketitle
\begin{abstract}
The problem of measurement in quantum mechanics is reanalyzed within a
general, strictly probabilistic framework (without reduction
postulate). Based on a novel comprehensive definition of measurement
the natural emergence of objective events is demonstrated and their
formal representation within quantum mechanics is obtained. In order
to be objective an event is required to be observable or readable in
at least two independent, mutually non-interfering ways with
necessarily agreeing results.  Consistency in spite of unrestricted
validity of reversibility of the evolution or the superposition
principle is demonstrated and the role played by state reduction, in a
properly defined restricted sense, is discussed. Some general
consequences are pointed out.
\\ \\ PACS numbers 
03.65.Bz, 
01.55+b,  
01.70+w   
\\ \\ \end{abstract}

\section{Introduction.}
Though quantum theory was formally completed over 60 years ago, its
conceptional foundations are still under debate [1-8]. Here a
resolution of the measurement problem \cite{Bell90,BLM} is presented,
reformulated in the following way: The essence of quantum mechanics
is, that it represents probabilities for the occurrence of events like
the ``click'' of a detector in a given arrangement. In order to be
complete, however, it must also be able to represent the individual
events themselves. It will be shown that it does indeed if
supplemented with a judicious and comprehensive definition of
measurement which allows one to express {\em objectivity} as {\em
independent verifiability} in concrete physical terms.\footnote{A more
condensed letter type version \cite{Sim91} was distribured earlier and
a short outline of the main definitions and results was given in
\cite{Sim92}.} {\em The basic postulate defining an event and
guaranteeing its objectivity, is, that it can be observed or read in
at least two independent, mutually non-interfering ways with
necessarily agreeing results.}

In order to focus on the essential physical concepts and simplify the
formulation, the analysis is performed in a general framework based
entirely on general ``common sense'' properties of probabilities for
the occurrence of events. Of course these properties are shared by, or
may be deduced from, conventional quantum
mechanics\footnote{Conceptually the present approach corresponds to
the ``minimal interpretation'' of ref. \cite{BLM}.} and it will in
fact be demonstrated explicitly, that the most simple minded quantum
mechanical models of measurements are covered by the results.  It is
emphasized, however, that the generality of the formulation is
essential for the completeness of the analysis of measurements which
must include any gathering and distribution of information on an
object, a fact not revealed by the commonly used models.

Events are conceptual objects with two values $e\in\{0, 1\}$ where
$e=1$ means that the event ``occurs'' and $e=0$ that it does not.
They are part of the interpretational language (metatheory) of quantum
theory and, unlike the probabilities of their occurrence, do not seem
a priory to have any counterpart within the mathematical structure of
the theory itself. However the present reanalysis of the concept and
description of measurements will in a natural way lead to mathematical
objects which exhibit, consistent with the requirements of quantum
theory, the objectivity properties characteristic of events as
postulated above. These are the mathematical objects representing
events in quantum theory.

It is emphasized that the purpose of this paper is to demonstrate the
natural emergence of objective events in quantum mechanics. It does
{\em not} aim at proving ``state reduction'' in the most general
sense, nor does it assume it. A {\em restricted} form of state
reduction, which is not at variance with the Schroedinger equation,
will be discussed, however, in particular with respect to the role it
plays for consistency of the definition and representation of events.

The remainder of this paper is organized as follows: In section 2 the
general probabilistic framework is outlined. Section 3 introduces the
complementary notions discrimination and interference with
superpositions defined for arbitrarily complicated states in a direct
probabilistic way not usually found in the literature. In section 4
measurements are discussed and defined mathematically. In section 5
discrimination is imposed and straight forward consequences are
stated. These lead to the central point of this work, discussed in
section 6, the natural emergence and mathematical representation of
objective events. In section 7 consistency of the so obtained
description of events with reversibility of the evolution and the role
played by state reduction etc. is discussed in detail. More
qualitative consequences and the conclusions are presented in the
remainder.

\section{General Probabilistic Framework.} 
To a given system belongs a set $\cal S$ of states, a set $\cal O$ of
observables and a function $(\cdot,\cdot):\ {\cal O \times S} \to
[0,1]$, i.e.
\begin{equation}\label{bounds}
0\le(A,X)\le 1\ \ \ \forall\ A\in{\cal O},\ X\in{\cal S},
\end{equation}
which is interpreted as probability $(A,X)=Prob(e=1)$ for the
occurrence of an event with value $e=1$ in a single observation or
{\em trial} on a state $X\in\cal S$ with a given observable $A\in\cal
O$.

In the conventional formulation of quantum theory $(A,X)=Tr[AX]$ where
$A$ and $X$ are hermitian operators on an appropriate Hilbert space
$\cal H$ obeying $0 \le A \le 1$ and $0\le X,\ Tr[X]=1$, respectively,
with $Tr$ denoting the trace over $\cal H$. If $X$ is a pure state
represented by a normed statevector $\varphi \in \cal H$, then
$X=|\varphi\rangle\langle\varphi|$ and
$(A,X)=\langle\varphi|A|\varphi\rangle$. {\em It is emphasized that
$X\in\cal S$ is a density operator or density matrix operating on
$\cal H$ and not an element of $\cal H$ itself and the term
``observable'' is used here only for the restricted class of positive
operators bounded by $1$ as indicated}.

$\cal S$ separates $\cal O$ i.e. $(A,X)=(B,X)\ \forall X\in\cal S$ 
iff $A=B$. To each $A \in \cal O$ a complement $\bar{A}\in\cal O$ 
exists which is defined by $(\bar{A},X)=1-(A,X)\ \forall X\in\cal S$ 
(and thus connected to $A$ by the replacement of the event $e$ by 
its negation $(1-e)$). Similarly a unit observable $I\in\cal O$ is 
defined by $(I,X)=1\ \forall X\in\cal S$ (i.e. setting $e=1$ 
independent of the state).

The set $\cal S$ is convex under classical (incoherent) mixing: To a 
given pair of states (density operators) $X_1,X_2\in \cal S$ there 
are mixed states $X=|c_1|^2 X_1 + |c_2|^2 X_2\,\in \cal S$ such 
that 
\begin{equation}\label{e10} 
(A,|c_1|^2X_1 + |c_2|^2X_2) = |c_1|^2 (A,X_1) + 
|c_2|^2 (A,X_2)
\end{equation}
where $\sum_i|c_i|^2=1$ (and $|c_i|^2 \ge 0$, of course) here and throughout this paper. A corresponding linearity property holds 
also for the observables in which case it can be extended to 
arbitrary real coefficients as long as the linear combination remains 
in $\cal O$. Obviously $\bar{A}=I-A$.

Of course all these statements are simple consequences of the quantum
mechanical formalism. It is emphasized, however, in particular in view
of the discussion of measurements below, that these general features
of $\cal S$, $\cal O$ and $(A,X)$ are indispensable for a consistent
probabilistic interpretation and can be deduced directly from it. They
apply also to classical mechanics with states represented by normed
density distributions on the appropriate phase space and observables
by corresponding measures. 

For mathematical definiteness and in order to emphasize where
appropriate that the set of observables admitted is not restricted in
any way appart from the above requirements, it is useful to define the
set $\widehat{\cal O}$ as the set of all {\em mathematically} possible
observables such that every (distinct) function $f: {\cal S}\to [0,1]$
obeying $f(|c_1|^2X_1+|c_2|^2X_2) = |c_1|^2f(X_1)+|c_2|^2f(X_2)$ is
represented by some $A\in\widehat{\cal O}$. Clearly ${\cal
O}\subseteq\widehat{\cal O}$ and $\widehat{\cal O}$ shares all the
properties of ${\cal O}$ given above.  $\widehat{\cal O}$ depends
solely on the set of states $\cal S$ itself and is not restricted in
any way by theoretical or ``feasibility'' considerations. Any
$A\in\widehat{\cal O}$ will be called an observable and, as a rule,
the reader ay assume ${\cal O}=\widehat{\cal O}$ without problems. Of
course $\widehat{\cal O}$ separates $\cal S$ i.e.  $(A,X)=(A,Y)\
\forall A\in\widehat{\cal O}$ iff $X=Y$ (which was not imposed on
$\cal O$).

\section{Discrimination and Interference.}
An observable $A$ {\em discriminates} between two states $X,
Y\in\cal S$ if $(A, X)=1$ and $(A,Y)=0$ or vice versa and thus if 
$|(A,X)-(A, Y)|=1$. If such an observable exists in $\widehat{\cal O}$,
$X$ and $Y$ are {\em orthogonal}.
Orthogonal pure states are represented by mutually 
orthogonal normed elements $\varphi,\psi$ in Hilbert space and an
observable which discriminates between them is given f.i. by
$A=|\varphi\rangle\langle\varphi|$. In general states are orthogonal
if the ranges of the density operators representing them are
orthogonal. An obvious physical example of a discriminating
observable is one representing a detector which discriminates between
a state concentrated in its fiducial volume (detection probability 1)
and one far away (detection probability 0).

More subtle concepts characteristic of quantum physics are 
superposition and interference. Here they are defined directly in 
terms of probabilities.

A state $X$ is a {\em (general) superposition} of two orthogonal 
states $X_1$ and $X_2$ with some fixed weights $|c_1|^2$ and 
$|c_2|^2$ if
\begin{eqnarray} \label{e15}
(A_1,X) =|c_2|^2 (A_1,X_2)\ \ &\forall A_1\in\widehat{\cal O} : (A_1,X_1)=0
\nonumber \\
(A_2,X) =|c_1|^2 (A_2,X_1)\ \ &\forall A_2\in\widehat{\cal O} : (A_2,X_2)=0.
\end{eqnarray}
The (convex) set of all states $X$ with this property is 
denoted by ${\cal S}(|c_1|^2 X_1,|c_2|^2 X_2)$. It obviously contains
the incoherent mixture $|c_1|^2 X_1+|c_2|^2 X_2$. 

Eq. (\ref{e15}) gives an operational definition which the reader is 
advised to visualize. It may easily be verified for the familiar 
{\em coherent} superposition of the form
$\varphi=c_1\varphi_1+c_2\varphi_2$ {\em viz.}
\begin{equation}\label{e18}
X=|\varphi\rangle\langle\varphi|=|c_1|^2|\varphi_1\rangle\langle\varphi_1|
+|c_2|^2|\varphi_2\rangle\langle\varphi_2| 
+c_1^*c_2|\varphi_2\rangle\langle\varphi_1| 
+c_2^*c_1|\varphi_1\rangle\langle\varphi_2|
\end{equation}
between two orthogonal pure states 
$|\varphi_i\rangle\langle\varphi_i|$ (since for $A\ge 0$ 
$\langle\psi|A|\psi\rangle=0$ implies 
$\langle\psi|A|\psi'\rangle=0\ \forall\psi'\in\cal H$). For a 
general state represented by an arbitrary density operator $X$ on 
$\cal H$, $X\in {\cal S}(|c_1|^2 X_1,|c_2|^2 X_2)$ if and only if 
there exist projection operators $P_1$ and $P_2$, $P_1+P_2=I$, such 
that $P_iXP_i=|c_i|^2X_i$. (A more complete treatement of
superpositions based on the probabilistic definition above will be
given elsewhere \cite{Sim93}.) 

Eq. (\ref{e15}) implies (replace $A_i$ by $\bar{A_i}$ where necessary) 
\begin{equation}\label{e16}
(A,X)=(A,|c_1|^2X_1+|c_2|^2X_2)
\ \ \forall X \in{\cal S}(|c_1|^2 X_1,|c_2|^2 X_2)
\end{equation}
if $(A,X_1)\in\{0,1\}$ \underline{or} $(A,X_2)\in\{0,1\}$ and thus 
in particular if $A$ discriminates between $X_1$ and $X_2$. 
Violation of eq. (\ref{e16}) for some observable $A$ represents an 
interference effect. An observable $A$ for which eq. (\ref{e16}) 
holds is {\em insensitive to interference} between $X_1$ and $X_2$. 
For the coherent superposition given in eq. (\ref{e18}) this has the 
familiar implication that $(A,X)$ depends on $|c_1|^2$ and $|c_2|^2$ 
only and not on their relative phase contained in $c_1^*c_2$, which 
is the case if and only if $\langle\varphi_1|A|\varphi_2\rangle=0$. 
In classical mechanics ${\cal S}(|c_1|^2 X_1,|c_2|^2 X_2)$ contains 
only $|c_1|^2 X_1 + |c_2|^2 X_2$ and eq. (\ref{e16}) therefor holds 
trivially.

\section{Measurements.}
The central feature of measurement on which the present analysis is
based, is the distribution of information about an object onto
several, at least two, different separately readable channels.  In
addition it may contain any type of manipulations, interaction with
external fields, passing through filters etc. in order to select
particular information on the object. Befor turning to the
mathematical representation and exact definition let me discuss its
physics in more detail:

In a measurement an object undergoes some interaction or interactions
with one or several other systems acting as probes or measuring
devices etc. such that afterwards there are several, at least two,
separated channels (identified in the following by greek upper
indices) consisting of different systems on which mutually
undisturbing (for the mathematical definition see [M4] below)
observations, using channel observables or {\em readings}
$A^\mu\in{\cal O}^\mu,\ \mu=1, 2,...$, may be performed in order to
obtain information on the object. Channels may consist of the original
object, the spin of a system or its spatial degrees of freedom, decay
products in a decay, photons, observers, bits in a computer, letters
in different copies of a paper, readers, friends and cats.... {\em Any
system to which information on the initial state of the system is
transferred which can be read by a separate observation is a channel
and any process which distributes such information onto different
channels constitutes a measurement.} The most abounding channels in
nature consist of photons.

In terms of individual events a measurement is characterized as
follows: In a single measurement or trial on a fixed initial state
$X\in\cal S$ of the object one can obtain
simultaneously\footnote{meaning for the same object or in the same
trial but not necessarily at exactly the same time!} a set of events
with values $\{e^\mu\}$ corresponding to fixed readings $A^\mu$, one
for each channel $\mu$ separately. These events can be combined to
give {\em coincidence events} with values $\prod e^\mu,\ \mu\in M$
where $M$ denotes a subset of channels. This will now be expressed in
terms of corresponding probabilities which can be represented in
quantum mechanics.

A given measurement is represented 
by a function $m(\{A^\mu,\ \mu\in M\};X)$, also denoted 
$m(A^1;X)$ or $m(A^1,A^2;X)$ etc., which depends on the readings 
$A^\mu,\ \mu\in M$ and the initial state $X\in\cal S$ of the object 
and is interpreted as the probability for such coincidences, 
$m(\{A^\mu,\mu\in M\};X)=Prob(\prod_{\mu\in M} e^\mu=1)$. It
of course depends on the interaction taking place during the
measurement and completely specifies the measurement including
dependence on initial states of probes or measuring devices etc. 

Channels may be grouped together into fewer combined channels. In this
way statements formulated below for two channels obtain general validity. A
coincidence observation between $A^\mu,\ \mu\in M$ is a (particular)
reading on the channel combined from all channels in $M$. 

The function $m$ extends to arbitrary observables 
$A^{tot}\in\widehat{\cal O}^{tot}$ 
on the total final state (including all channels) of the measurement
and has the following \mbox{{\em defining properties} [M0]--[M5]:}\\
{[M0]} $m(;X)=1\ \forall X\in\cal S$ for the empty set $M=\emptyset$
of channel readings i.e. no reading at all.\\
{[M1]} $0\le m(A^{tot};X)\le 1$ for all $X\in\cal S$ and
       $A^{tot}\in\widehat{\cal O}^{tot}$.\\ 
{[M2]} Convex linearity in $X\in\cal S$ as for $(A,X)$ in eq. (\ref{e10}).\\
{[M3]} Global linearity for arbitrary final state observables 
corresponding to the discussion after eq. (\ref{e10}).\\
{[M4]} Separability or mutual non-disturbance of readings of 
different channels expressed by 
\begin{equation}\label{e60}
m(A^\mu,A^\nu;X)+m(A^\mu,\bar{A^\nu};X)=m(A^\mu;X) 
\end{equation}
($\mu\ne\nu$) for all $A^{\mu,\nu}\in{\cal O}^{\mu,\nu}$
independentof\footnote{For {\em fixed} $A^\nu$ this just {\em defines}
$m(A^\mu;X)$ as the marginal over the outcomes of $A^\nu$. The
non-trivial part of the requirement is that this marginal is {\em
independent of the choice of $A^\nu$.}} $A^\nu$, and correspondingly
for an arbitrary number of channels i.e. $A^\mu$ and $A^\nu$ replaced
by observations on any two disjoint sets of channels. Here the l.h.s.
means that the information corresponding to $A^\nu$ is ignored though
it has been obtained, and the r.h.s that no observation of channel
$\nu$ is performed at all.\\
{[M5]} Linearity in the readings $A^\mu\in{\cal O}^\mu$ of each
channel $\mu$ separately.\\
{\em This defines measurements mathematically. Every function 
$m: {\cal S}\times {\cal O}^1\times {\cal O}^2\times\cdots\to[0,1]$
with these properties represents a possible measurement.}

[M1]--[M3] are dictated by the probability interpretation as discussed
for $(\cdot,\cdot)$ in sec. 2.  The central property used here is the
separability condition [M4] and thus eq. (\ref{e60}). It constrains the
function $m$ and thus the arrangements qualifying for a measurement.
As a rule, though not a strict one, it requires the reading of
different channels to take place or be restricted to separate space
regions. Eq. (\ref{e60}) implies
\begin{equation}\label{e61} 
m(A^\mu,A^\nu;X) \le m(A^\mu;X).
\end{equation}
[M5] is not independent of [M3] and [M4], but the connection is rather
subtle. For simplicity [M5] is therefor just stated here as an
additional condition.

The standard measurement in nature is scattering of photons from 
some source (e.g. the sun) on a (usually macroscopic) object (e.g. the moon). 
Observation of different photons (looking at the moon) in different 
space regions are mutually non-disturbing (unless e.g. one person 
stands in front of the other).

The ``standard'' measurement in quantum mechanics discussions is 
the certainly oversimplified model based on transitions
\begin{equation}\label{e20}
\varphi_i \otimes \psi_o \to \varphi_i \otimes \psi_i 
\end{equation}
with 
$\langle\varphi_i|\varphi_j\rangle=\langle\psi_i|\psi_j\rangle=\delta_{ij}$.
For arbitrary normed $\varphi = \sum_i c_i\varphi_i$ this implies 
\begin{equation}\label{e22}
m(A^1,A^2;|\varphi\rangle\langle\varphi|)
\nonumber\\  
=\sum_{ij}c_ic^*_j\langle\varphi_j\otimes\psi_j|A^1\otimes 
A^2|\varphi_i\otimes\psi_i\rangle
=\sum_{ij}c_ic^*_j\langle\varphi_j|A^1|\varphi_i\rangle
\langle\psi_j|A^2|\psi_i\rangle
\end{equation}
and if only the second channel is read f.i. 
\begin{equation}\label{e23}
m(A^2;|\varphi\rangle\langle\varphi|)  
=\sum_{ij}c_ic_j^*\langle\varphi_j|\varphi_i\rangle
\langle\psi_j|A^2|\psi_i\rangle 
=\sum_{i}|c_i|^2\langle\psi_i|A^2|\psi_i\rangle.
\end{equation}
Of course the $\varphi$ and $\psi$ refer to the two channels of this
measurement the first one of which is the same as the object itself in
this case. The model in eqs. (\ref{e20}-\ref{e23}) may easily be
generalized to an arbitrary number of channels by replacing Hilbert
space, states and observables of channel two by corresponding tensor
products for an arbitrary number of channels. In addition, instead of
the simple model evolution (\ref{e20}) an arbitrary unitary transition
\begin{equation}\label{e25}
X\otimes X_m \to S(X\otimes X_m)S^\dagger
\end{equation}
may be adopted where $S$ is a unitary evolution or scattering operator
and $X$ and $X_m$ are statistical operators describing object and 
initial state of the measuring device, respectively. 

[M1]--[M5] are easily verified in all these cases. In fact [M1]
follows from the tensor 
product representation of the initial state and [M2]--[M5] from the 
representation of coincidence readings by tensor products 
$A^\mu\otimes A^\nu\otimes\cdots$ of channel observables with the
standard rule that the partial trace is to be taken over the Hilbert
spaces of all channels which are not observed. In particular eq.
(\ref{e60}) follows from $A^\mu\otimes\bar{A}^\nu =
A^\mu\otimes(I^\nu-A^\nu) = A^\mu\otimes I^\nu-A^\mu\otimes A^\nu$.
It should be emphasized, however, that [M1]--[M5] are the primary
physical requirements in terms of probabilities {\em entailing or at
least permitting} the tensor product representation rather than being
a consequence of it. In particular 
[M4] is a necessary condition in order for the tensor product
representation to be applicable. It is far from being a trivial formal
feature.

For given $\{A^\mu,\mu\in M\}$ $m$ corresponds to an observable
$F[\{A^\mu,\mu\in M\}]\in\widehat{\cal O}$ such that 
\begin{equation}\label{e50} 
m(\{A^\mu,\mu\in M\};X) = (F[\{A^\mu,\mu\in M\}],X).
\end{equation}

Measurements can be composed, the number of channels enlarged, 
and information transmitted to new ones by performing measurements 
on channels of previous ones. Upon reading the ``reader'', whether 
photon or human being, herself becomes a channel. 
There is no need for a formal implementation of this; in the sequel
only the fact will be used that the final result of arbitrary such
manipulations leads to a measurement describable by a
function $m$ with the properties listed.

As introduced here, a measurement is a {\em means of observation} of
the (initial) state $X$ of an object. However, it also {\em prepares}
a state $X^\nu$ of any (or all) of its (final) channels, in particular
the final state of the object itself if considered as a channel, such
that
\begin{equation}\label{e28}
(A^\nu,X^\nu)=m(A^\nu;X)
\end{equation}
for arbitrary $A^\nu$ but fixed $X$ or, with additional
selection\footnote{i.e. restriction to a subensemble of the total
ensemble under consideration when the measurement is performed
repeatedly.} based on a fixed reading $A^\mu$ of a ``selection
channel'' $\mu\ne\nu$,
\begin{equation}\label{e29}
(A^\nu,X^\nu)
=\frac{m(A^\nu,A^\mu;X)}{m(A^\mu;X)}
\end{equation}
(= conditional probability for $e^\nu=1$ given $e^\mu=1$). 
For the case of eq. (\ref{e22}) the states $X^1$ of the object after 
the measurement corresponding to eqs. (\ref{e28}) and (\ref{e29}) are,
respectively, 
$X^1=\sum_i|c_i|^2|\varphi_i\rangle\langle\varphi_i|$
and
$X^1=|\varphi_i\rangle\langle\varphi_i|\ \ if\ \ 
A^2=|\psi_i\rangle\langle\psi_i|$.
(The absence of interference terms (containing $c_i^*c_j$, $i\ne 
j$) in these two equations as well as in eq. (\ref{e23}) is due to the 
fact that this measurement is discriminating as discussed in detail in the 
next section.)

Though closely related the application of measurements to observation
and to preparation must be kept apart. 

\section{Discriminating Measurements.}
Throughout the following $X_i$ refers to states of the object on which
the measurement is performed, $X_i\in \cal S$, and $A^\mu$ to a
reading of channel $\mu$ of the measurement i.e. $A^\mu\in{\cal
O}^\mu$.

A channel $\mu$ of a measurement {\em discriminates} between two 
states $X_1$ and $X_2$ (requiring $X_1$ and $X_2$ to be orthogonal) 
if reading of that channel alone, without further observation, 
allows one to discriminate between $X_1$ and $X_2$. A corresponding 
reading $A^\mu$ {\em discriminates $X_1$ against $X_2$} if
\begin{equation}\label{e70} 
m(A^\mu;X_i)=\delta_{i1}.
\end{equation}
Then $A^\mu$ and $\bar{A^\mu}$ both discriminate between $X_1$ and
$X_2$. 

Eq. (\ref{e22}) represents a measurement which discriminates between
$|\varphi_i\rangle\langle\varphi_i|$ and
$|\varphi_j\rangle\langle\varphi_j|$ due to the requirement
$\langle\varphi_i|\varphi_j\rangle=\langle\psi_i|\psi_j\rangle=\delta_{ij}$
imposed, and $A^1=|\varphi_i\rangle\langle\varphi_i|$ or
\mbox{$A^2=|\psi_i\rangle\langle\psi_i|$} both discriminate
$|\varphi_i\rangle\langle\varphi_i|$ against
$|\varphi_j\rangle\langle\varphi_j|$ for $j\ne i$. In this case, as in
many measurements, the object itself represents a discriminating
channel. The fact that the $X_i$ are unchanged by the measurement,
however, is an idealization (measurement of the first kind) which is
too restrictive and not imposed in general.  But this simple model is
admissible and the following central results are easily verifyed for
it explicitly.

Throughout the following ``discriminating'' refers to two 
orthogonal states $X_1$ and $X_2$ even if they are not mentioned and 
a discriminating measurement has at least two discriminating 
channels. The term ``sensitive to interference'' will be applied to 
measurements and channels correspondingly.

\begin{thm}[Probability] \label{t1}
If a reading $A^\mu$ of a channel $\mu$ of a measurement 
discriminates $X_1$ against $X_2$ then 
\begin{equation}\label{e75}
m(A^\mu;X)=|c_1|^2\ \ \forall X\in{\cal S}(|c_1|^2X_1,
|c_2|^2X_2).  
\end{equation}
\end{thm}
This familiar rule follows from eqs. (\ref{e50}) and (\ref{e15}).
\qed
\begin{thm}[State reduction] \label{t2}
Consider a measurement with a reading $A^\mu$ which discriminates
between two states $X_1$ and $X_2$ and let $X\in{\cal
S}(|c_1|^2X_1,|c_2|^2X_2)$ for some $c_i,\ |c_1|^2+|c_2|^2=1,$ be any
superposition between the $X_i$. Then for arbitrary reading $A^\nu$,
$\nu\ne\mu$, of any other channel (or combination of channels),
whether discriminating or not,
\begin{equation}\label{e81}
m(A^\nu,A^\mu;X)
=m(A^\nu,A^\mu;|c_1|^2X_1+|c_2|^2X_2)
\end{equation}
\begin{equation}\label{e82}
m(A^\nu;X)=m(A^\nu;|c_1|^2X_1+|c_2|^2X_2)
=|c_1|^2m(A^\nu;X_1)+|c_2|^2m(A^\nu;X_2).
\end{equation}
Thus $m(A^\nu,A^\mu;X)$ and $m(A^\nu;X)$ are both insensitive to
interference between the $X_i$ and therefor any observation on any
combination of channels (here collectively represented by $\nu$) which
excludes some discriminating channel (here $\mu$) is insensitive to
interference between them. 
\end{thm}
{\em Proof: }
Due to eqs. (\ref{e50}) and (\ref{e61}), 
eq. (\ref{e81}) follows, as eq. (\ref{e16}), from eq. (\ref{e15}). 
Adding to eq. (\ref{e81}) the same relation with $A^\mu$ replaced by 
$\bar{A^\mu}$ and using eq. (\ref{e60}) one obtains eq. (\ref{e82}) (with
linear expansion based on [M2]).
\qed
For the two-channel model of eq (\ref{e22}) with $\nu=1$, $\mu=2$ and 
$X_i=|\varphi_i\rangle\langle\varphi_i|$, eq. (\ref{e81}) is verified with
$A^2=|\psi_1\rangle\langle\psi_1|$ or $|\psi_2\rangle\langle\psi_2|$
and eq. (\ref{e82}) corresponds to eq. (\ref{e23}). 
\begin{thm}[Objectivity] {\em\cite{Sim92}} \label{t3} 
Consider a measurement with channels $\mu\in M$ which discriminate
between two states $X_1$ and $X_2$ and choose readings $A^\mu$, $\mu
\in M$, which discriminate $X_1$ against $X_2$ according to eq.
(\ref{e70}). If $X$ is any superposition between the $X_i$, i.e
$X\in{\cal S}(|c_1|^2X_1,|c_2|^2X_2)$ for some $c_i$ with
$|c_1|^2+|c_2|^2=1$, then for arbitrary $\mu,\nu\in M$, $\mu\ne\nu$ 
\begin{equation}\label{e92}
m(A^\mu,\bar{A^\nu};X)=m(\bar{A^\mu},A^\nu;X)=0
\end{equation}
or, equivalently,
\begin{equation}\label{e93}
m(A^\mu,A^\nu;X)=m(A^\mu;X)=m(A^\nu;X).
\end{equation}
\end{thm}
{\em Proof: }
By supposition $m(A^\rho;X_2)=m(\bar{A^\rho};X_1)=0$ 
$\forall \rho \in M$ implying with eq. (\ref{e61}) that eq. 
(\ref{e92}) holds for $X=X_i$. It then holds for all $X\in{\cal 
S}(|c_1|^2X_1, |c_2|^2X_2)$ due to eqs. (\ref{e15}) and 
(\ref{e50}). Equivalence between eqs. (\ref{e92}) and (\ref{e93}) is due
to eq. (\ref{e60}).
\qed

These are the central theorems for the analysis of measurements.  It
is emphasized that they are based solely on the defining mathematical
properties of the functions $m(\cdot;\cdot)$ and $(\cdot,\cdot)$
without explicit reference to their interpretation in terms of events
(which of course motivated these properties). Compatibility with all
conventional rules of quantum mechanics, including the linearity of
the law of motion, is manifest from eqs. (\ref{e20}--\ref{e25}).

It is important to note the general structure of the three theorems:
The premises involve only the response of each channel separately to
the two states $X_1$ and $X_2$. No assumptions are made about
superpositions between those states nor about correlations among the
different channels. The behaviour for arbitrary superpositions as well
as the crucial corelletions between different channels (for all such
superpositions) are then obtained as mathematical consequences.
Experimental verification of the premises is therefor possible without
preparing superpositions between $X_1$ and $X_2$ (which in many cases
of interest could be very difficult).

\section{Probability and Objective Events.}
For $X\ne X_i$ the actual value of an event is not predicted, only 
its probability as given by theorem \ref{t1}.

However, for a given trial the probabilistic element is eliminated for
all $X\in{\cal S}(|c_1|^2X_1,|c_2|^2X_2)$ according to theorem
\ref{t3}: Under the conditions of theorem \ref{t3} the readings of different 
discriminating channels necessarily agree. This constitutes an
objective event stored in the collection of discriminating channels of
the measurement which can be read on different channels independently
with zero probability of disagreement as shown by eq.  (\ref{e92})
(even long time after, if the channel is not destroyed implying
usually that separate copies of the channels are involved for
different trials). Theorem \ref{t3} thus has the
\begin{cor}[Objective Events] {\em\cite{Sim92}} \label{c3}
Under the condition of theorem \ref{t3} the function $m(\cdots;X)$
represents an event with an objective value $e\in\{0,1\}$ for each
trial: A measurement which discriminates between two states $X_1$ and
$X_2$ performed on any superposition $X$ between them produces for
each trial an (objective) event for an observable $A$ which
discriminates between $X_1$ and $X_2$. The value of this event can be
obtained or read from each discriminating channel separately.
\end{cor}
This is the central result of this paper. While quantum mechanics and
in particular the function $m$ in general {\em predicts} only the {\em
probabilities} of events and not their values, it nevertheless
consistently {\em describes} the events themselves. At least two
discriminating channels are required for objectivity, i.e. independent
verifiability, but in principle also enough, though this is a rather
unnatural model case. The separability condition [M4] is the basis for
the required independence of the readings of different channels.  It
is emphasized that a corresponding separability or independence
condition cannot hold in general for observables $A^\mu$ and $A^\nu$
acting on the {\em same} object or channel unless all observables are
compatible (commute).

The states $X_i$ between which a measurement discriminates\footnote{It
should be clear, that, depending on the reading, a measurement may
discriminate between many different mutually orthogonal states. In the
extreme case, it can in principle discriminate between all members of
a family of mutually orthogonal states
$|\varphi_i\rangle\langle\varphi_i|$. If $\{\varphi_i\}$ forms a basis
of the Hilbert space in question, such a a measurement
corresponds to a so-called ``complete set of commuting observables''
(to be distinguished from a separating set of observables in the sense
discussed at the end of section 2).} define what the measurement is
good for or ``what it measures''. This condition is formulated for
each channel separately defining what infiormation can be obtained
from that channel. The correlation between different channels is not
imposed but obtained as a consequence.

Obviously one has to know how to read the channels correctly according
to eq. (\ref{e70}) as one has to know the meaning of the dial of any
instrument one uses\footnote{as one has to learn the meaning of 0 and
1 in the first place}. But this condition involves $X_1$ and $X_2$
only and not their superpositions: it is not necessary to prepare
superpositions between the $X_i$ in order to learn how to read the
result of the measurement.

{\em Remark:}
Measurements performed on microscopic objects often do not discriminate
between any two states of the object itself either since the interaction of
the object with other matter is too weak, as e.g. for neutrinos, or
bacause no analyzer with optimal efficiency is available otherwise, as
e.g. for the spin of a particle, or for various other reasons. Such
measurements 
still lead to objective events if at some intermediate stage they
proceede over some trigger state or states (of the object itself or
some other trigger system like an atom which can be ionized by the
object) which are discriminated by the subsequent measurement. The
analysis of discriminating measurements then applies to the
measurement performed on these trigger states. Discrimination at some
stage is obviously a necessary prerequisite for the emergence of an event
(since it must be possible to discriminate between $e=1$ and $e=0$).
The trigger provides the common cause for the complete correlation
between different channels required by the definition of the event.

\section{State Reduction, its Limitations, and Consistency.} 

Remains to discuss in more detail consistency in view of the
reversibility or linearity (on Hilbert space) of the basic law of
motion (superposition principle) which implies that coherence of
superpositions is strictly conserved in the evolution. Thus, at least
to the extent that object and measuring device, including all
channels, together can be considered to be an isolated system,
observations sesitive to interference between two states must be
possible in principle also after a measurement which discriminates
between them.  The present definitions and analysis do not preclude
this as may be verified in the explicit example represented by eq.
(\ref{e22}). But contrary to widespread opinion this does not lead to
any contradiction, as shown by theorem \ref{t2}. Though it does not
completely preclude them, theorem
\ref{t2} crucially restricts observations sensitive to interference
between two states $X_1$ and $X_2$ after a measurement which
discriminates between them.  In fact it requires that an observation
which is sensitive to interference {\em must} include all potentially
discriminating channels of the measurement in such a way, that no
discriminating channel survives and no discriminating reading is
possible anymore (for the same trial) which in turn implies that the
event is completely erased (also from the memory of eventual readers).
For instance in eq. (\ref{e22}) both
$\langle\varphi_1|A^1|\varphi_2\rangle$ and
$\langle\psi_1|A^2|\psi_2\rangle$ are required to be non-zero for
sensitivity to interference and if there are more channels the same is
required for all of them. Let me make this clear: {\em It is {\bf not}
possible to first read the result of the discriminating measurement
somehow (also not from a channel with ``macroscopic'' or ``classical''
properties) and in addition, in the same trial, perform an observation
on the original channels of the measurement which is sensitive to
interference.} Indeed, the carrier of the read information itself is
now a discriminating channel and according to theorem
\ref{t2} therefor has to be included in any observation sensitive to
interference.  {\em All} (possible) discriminating channels have to be
included and not only those explicit in some simplified model. {\em The
crucial feature of theorem \ref{t2} is that it precludes coexistence of
discriminating channels with channels sensitive to interference.}

Thus theorem \ref{t2} is sufficient to guarantee consistency of the
definition of objective events in spite of the reversibility of the
law of motion or the superposition principle, since it implies that it
is principally impossible to have an objective event in the sense of
corollary \ref{c3} and at the same time, i.e. in the same trial, obtain
information on, or be sensitive to, interference between the two
states discriminated. There is no need whatsoever from a basic point
of view to prove or postulate the absolute impossibility of obtaining
interference effects after a discriminating measurement and thus no
need to restrict the applicability of the superposition principle or
modify the law of motion \cite{Weber} in order to introduce
irreversibility.

At face value this argument may look difficult. But actually what is
behind it is very simple: {\em One just cannot violate the defining
equations (\ref{e15}) of superpositions however complicated one chooses
the arrangement for an observation to be }(cf. the proof of theorem \ref{t2}). 

The distinction between an event and a probability is crucial: The
mere existence of an observable sensitive to interference does not imply
that a corresponding event has been generated. The information
corresponding to an event must be distributed onto different channels.
Otherwise, as mentioned befor, independent readability respecting
condition [M4] is not guaranteed.

It is emphasized that rather than requiring that a measurement leads
to a ``registered'' result which can be distributed, any such
distribution of information is, and for consistency must be, included
in the analysis. Moreover, it is not sufficient to stipulate only that
information {\bf can} be distributed; quantum mechanics requires that
it {\bf must be} distributed. The difference is crucial as exemplified
by the case of a spin system: (discriminating) information on the spin
in any direction one choses {\bf can} be distributed at any time (by a
corresponding measurement) but only the information on the spin in
{\bf one} fixed direction can {\bf be} distributed (by theorem
\ref{t2}).

Though strictly speaking not relevant for the question of objectivity
and consistency, some additional features of theorem \ref{t2} are
noteworthy:

In spite of its limited scope, theorem \ref{t2}, in conjunction with the
restrictions due to relativistic kinematics, does, under appropriate
conditions, entail absolute impossibility of obtaining interference
effects ``after the fact'' in a certain sense:
\begin{cor}[Irreversible state reduction] \label{c4}
If in a measurement some channel which discriminates between two 
states consists of photons which escape into free space, then no 
subsequent observation can be sensitive to interference between 
these two states unless the equipment which has to intercept the 
photons in order to detect interference effects (theorem \ref{t2}) is 
installed and activated in beforehand. Otherwise the loss of 
coherence is irreversible. 
\end{cor} 

Of course theorem \ref{t2} also shows the in general tremendous {\em
practical} difference between a discriminating reading (involving only
one discriminating channel) and an observation sensitive to
interference (which has to involve all discriminating channels).
Clearly, to provide simple access to information is the purpose of
measurement. 

Finally theorem \ref{t2} shows that the states $X_i$ between which a
measurement can discriminate objectively are determined by the
measurement itself i.e. the interaction taking place in it's course.
They are fixed once at least two discriminating channels are
separated. This is where and when the ominous reduction of the state
of the separate channels, without selection according to eq.
(\ref{e29}), takes place.

\section{Further Consequences and Discussion.}
So far discussion centered around discrimination between two states 
only. However, using an appropriate measurement with many channels 
and judicious readings $A^\mu$ for different channels one can 
``filter out'' the state of an object in an individual 
trial\footnote{This actually {\em defines} the term simultaneous 
measurement. For fixed readings $A^\mu$ the coincidences together 
with complementation generate a boolean structure (coincidence 
logic). (Note that if the same reading is used in different 
coincidences the information has to be split into different channels 
by a corresponding discriminating measurement.)}  f.i. if for each 
$X_i$ in a set of mutually orthogonal states a reading in some 
channel is chosen which discriminates this state against all the 
others. Schematically this is how the eye determines the 
(approximate) location of the moon.

Contrary to widespread opinion no macroscopic features (environment
induced or otherwise) of measurement devices etc. have to be {\em
invoked} in order to implement objectivity. On the othe hand, the
present analysis {\em shows} (qualitatively) why macroscopic objects
(e.g. the moon) can be described individually and are found always in
macroscopically localized states and never in superpositions between
such states \cite{Sim78}: macroscopically different states can be
discriminated by observing the light permanently scattered on them
within their natural surrounding. Therefore, coherence between them
can not prevail according to theorem
\ref{t2} and eq. (\ref{e28}) (see also corollary \ref{c4}) and different
observers will necessarily agree on the state they see according to
theorem \ref{t3} and the discussion thereafter. It should be noted,
that light scattered on a macroscopic object cannot be sensitive to
{\em interference} between macroscopically different states $X_i$.
Otherwise, such light, if scattered on the $X_i$ themselves, would
have to intermix them in such a way as to render subsequent
discrimination between them impossible (theorem \ref{t2}). Obviously
this is not the case; no superpositions between the $X_i$ are needed
to verify this experimentally. In fact, if it would be the case, it
would render macroscopically localized states unstable under the
influence of light (and thus make a real mess out of our macroscopic
world). Thus outside influence singles out and {\em defines}
macroscopic states \cite{Sim78}.

If a channel of a measurement is macroscopic (a ``pointer'')
discrimination between two states $X_1$ and $X_2$ of the object must
of course rely on discrimination between macroscopically different
states of the ``pointer''. If it would rely on interference between
macroscopically different states of the ``pointer'' the information
would be lost\footnote{See also
\cite{Zurek}. According to theorem \ref{t2} and the discussion at the
end of section 7 it is {\em not} so, however, that some other
information, involving interference between $X_1$ and $X_2$, would be
obtainable instead by reading the pointer.}. Clearly macroscopic
``pointers'' play an important role since their (macroscopically
different) states attain objectivity in a natural way ``all by
themselves''. But this is not a prerequisite for objectivation, only
its most convenient and natural realization. In fact it should be
mentioned that the usual
analyzes based on macroscopic features of pointers do not actually
{\em prove}, but {\em assume} objectivity based on classical physics
where, of course, this assumption is unproblematic.\footnote{This is
true also for attempts at objectivation based on hidden
variables etc.}

The analysis presented here is based on a general probabilistic
formulation. Simple and arbitrarily complicated states are treated on
the same footing. Besides basic non-relativistic quantum mechanics and
simple minded models based on it, it includes relativistic field
theory, $C^*$-algebraic or whatever approach one prefers. On the other
hand, whether one wants to describe some channel and readings quantum
mechanically, though never excluded, can be left open. (If a channel
consists of your friend ask her the result and never mind about
operators.) 

The definition of measurements and objectivity rests on separability
as expressed in a probabilistic way in eq. (\ref{e60}). Actually, as a
concept, separability is indispensable in many other respects and of
course usually just tacitly assumed (knowingly or not). In fact the
basic definitions of states of an object make use of the assumption
that the only information transfered between preparation and
observation of the state is contained in the state of the object
itself. It is emphasized that Bell type inequalities \cite{Bell64} can
not be deduced within a purely probabilistic framework as used here
unless additional assumptions are made which, in particular in view of
the present results, are not compelling.

Several aspects of this analysis could only be touched 
upon superficially or not at all here. I plan to come back to them in 
more detail elsewhere.

\section{Conclusions.}
In conclusion it has been shown that, supplemented with appropriate
definitions of measurements and objectivity, quantum mechanics, with
only its minimal \cite{BLM}, probabilistic interpretation, {\em can}
describe individual objective events and not merely the probability of
their occurrence and that ``the moon is objectively there, even if
nobody looks'' \cite{Einstein}. Mathematical objects representing
objective events in quantum mechanics have been identified. The
analysis is based on mathematically rigorous results (theorem
\ref{t1}-\ref{t3}) with the central consequence stated in corollary
\ref{c3} and consistency discussed in detail in setion 7.  No random
phase assumption \cite{vanKampen}, explicit macroscopic feature
\cite{Jauch}, last observer \cite{vonNeumann}, modification of the
basic linear law of motion (Schroedinger equation) \cite{Weber} nor
any (other) reference to FAPP -- For All Practical Purposes
\cite{Bell90} -- is required. The main input is the direct expression
of superpositions in terms of probabilities given in section 3 and the
generality of the concept of measurement which applies to any
distribution of information about the object and to any possible
carrier of such information.

Crucial is of course the notion of objectivity and objective event
used in the present analysis. Its implementation into quantum physics
requires that information is distributed onto different channels from
which it can be read independently.  Rather than requiring that a
measurement leads to a ``registered'' result which {\bf can} be
distributed nondestructively, any such distribution of information is,
and for consistency must be, included in the analysis based on the
comprehensif mathematical definition of measurement presented here.


\begin{thebibliography}{99}

\bibitem{Wheeler}
{Quantum Theory and Measurement, edited by J. A. 
Wheeler and W. H. Zurek (Princeton University Press, Princeton, 
1983).}

\bibitem{Esp}
{B. d'Espagnat, Conceptual Foundation of Quantum Mechanics 
(Benjamin, Reading, MA, 1976), 2nd ed.}

\bibitem{Bell87}
{J. S. Bell, Speakable and Unspeakable in Quantum 
Mechanics, (Cambridge University Press, Cambridge, 1987).}

\bibitem{Shimony}
{A. Shimony, Scientific American 258 (1988) 36.}

\bibitem{62years}
{Sixty-Two Years of Uncertainty, Proceedings of the NATO Advanced 
Study Institute, Erice 1989, A. I. Miller ed. (Plenum Press, New 
York 1990).}

\bibitem{Bell90}
{J. S. Bell, Against measurement, ref. \cite{62years} p. 17, 
reprinted in Physics World, August 1990, p. 33.}

\bibitem{BLM}
{P. Busch, P. J. Lahti and P. Mittelstaedt, The Quantum Theory of
Measurement, Lecture Notes in Physics m2 (Springer, Berlin etc. 1991).}

\bibitem{Gottfried 1991}
{K. Gottfried, Physics World, October 1991, p. 34.}

\bibitem{Sim91}
{M. Simonius, Measurement in Quantum Mechanics: From Probabilities to
Objective Events, Preprint, March 1991, unpublished} 

\bibitem{Sim92}{M. Simonius, Helv. Phys. Acta 65 (1992) 884.}

\bibitem{Sim93}{M. Simonius, in preparation}

\bibitem{Weber}
{G. C. Ghirardi, A. Rimini and T. Weber, Phys. Rev. D 34 (1986) 470.}

\bibitem{Zurek}
{W. H. Zurek Phys. Rev. D 24 (1981) 1516 and D 26 (1982) 1862.}

\bibitem{Sim78}{M. Simonius, Phys. Rev. Letters 40 (1978) 980.}

\bibitem{Bell64}{J. S. Bell, Physics 1 (1964) 195.}

\bibitem{Einstein}{A. Einstein, Philos. Sci. 1 (1934) 162, A. 
Pais, Rev. Mod. Phys. 51 (1979) 863, p.907.}

\bibitem{vanKampen}{N. G. van Kampen, Physica A 153 (1988) 97.}

\bibitem{Jauch}{J. M. Jauch, Helv. Phys. Acta 37 (1964) 311.}

\bibitem{vonNeumann}{J. von Neumann, Mathematische Grundlagen der
Quanten Mechanik (Springer, Berlin 1932, reprinted 1968).}

\end{thebibliography}
\end{document}